\documentclass[10pt,letterpaper]{article}

\usepackage{amsmath}
\usepackage{amssymb}
\usepackage[dvips]{graphicx}
\usepackage{psfrag}
\usepackage{amsthm}

\newcommand{\diff}[2]{\frac{d #1}{d #2}}
\newcommand{\pdiff}[2]{\frac{\partial #1}{\partial #2}}
\newcommand{\bra}[1]{\langle #1 |}
\newcommand{\ket}[1]{| #1 \rangle}

\newcommand{\bs}{\boldsymbol}
\newcommand{\eps}{\varepsilon}
\newcommand{\atanh}{\textrm{arctanh}}

\theoremstyle{definition} 

\title{The Two Dimensional Kondo Model with Rashba Spin-Orbit Coupling}
\author{Justin Malecki\footnote{jjmaleck@physics.ubc.ca} \\ 
\textit{Department of Physics and Astronomy} \\  \textit{University of British Columbia} \\ \textit{Vancouver, Canada}}
\date{}

\begin{document}

\maketitle

\begin{abstract}
We investigate the effect that Rashba spin-orbit coupling has on the low energy behaviour of a two dimensional magnetic impurity system.  It is shown that the Kondo effect, the screening of the magnetic impurity at temperatures $T < T_K$, is robust against such spin-orbit coupling, despite the fact that the spin of the conduction electrons is no longer a conserved quantity.  A proposal is made for how the spin-orbit coupling may change the value of the Kondo temperature $T_K$ in such systems and the prospects of measuring this change are discussed.  We conclude that many of the assumptions made in our analysis invalidate our results as applied to recent experiments in semi-conductor quantum dots but may apply to measurements made with magnetic atoms placed on metallic surfaces.

\end{abstract}

\section{Introduction}

The physical behaviour of a single magnetic impurity interacting with a large number of conduction electrons has been, and continues to be, a fascinating subject of inquiry for over four decades.  The emergence of a strong correlation between the impurity and the conduction electrons at low temperatures, first predicted by Jun Kondo~\cite{kondo}, gives rise to the well-known Kondo effect wherein the impurity is effectively screened by forming a singlet with the surrounding electrons.  Much still remains to be understood regarding strongly correlated quantum many-body physics.  The Kondo effect, as a well understood example of a strongly correlated phenomenon, has proven to be a fertile topic for further study in this important field~\cite{hewson}.

In recent years, the study of these magnetic impurity systems has seen a resurgence due to the great advances in experimental technology that allow one to construct, control, and manipulate objects on the nanometer scale.  In particular, the Kondo effect has been observed in numerous experiments involving quantum dots constructed in semiconductor heterostructures~\cite{goldhaber}, as well as in systems where a magnetic atom is placed on the surface of metal~\cite{stm}.

Both of these systems involve a local moment, either that of an odd number of localized electrons (as in the quantum dot) or of a partially filled outer shell of an atom  interacting with mobile electrons confined to move in two dimensions.  The confining potential breaks the inversion symmetry that is manifest in the conduction electrons and gives rise to a coupling between their spin and momentum~\cite{rashba}.  Such spin-orbit coupling (coined as Rashba spin-orbit coupling to differentiate it from coupling arising from other mechanisms which may break the inversion symmetry) has been analyzed theoretically and observed experimentally in both semiconductor quantum wells~\cite{winkler,sosemiconductorexp} and in metallic surface states~\cite{lashell,surfacerashba,surfaceexp}.

The presence of such a spin-orbit coupling means that the $SU(2)$ symmetry of the conductions electrons is no longer present and that the spin of the electrons is no longer an eigenstate of the system.  Given this fact, it seems, at first, remarkable that the Kondo effect, which depends crucially on the magnetic exchange interaction between the local moment and the spin of the conduction electrons, is observed at all in such two dimensional systems.  This issue was first examined in~\cite{mw} where it was shown that the presence of time reversal symmetry in the Kondo interaction term preserves the spin coherence of the electron propagator.  That is, although spin is no longer a ``good'' quantum number, the scattering trajectories of the electron that would destroy its spin coherence in multiple interactions with the impurity are exactly cancelled by the time reversed paths.  While such a powerful argument may explain why the Kondo effect survives the apparent breaking of spin coherence, it is not precise enough to determine the change in the low energy behaviour of such impurity systems in the presence of a spin-orbit coupling in general.  A more detailed calculation is required in order to gain such information.

There have been several studies that have investigated the role of spin-orbit effects in specific semiconductor mesoscopic devices where the Kondo effect is manifest~\cite{sun,simonin,lopez,dingdong}.  In these studies, the spin-orbit interaction either takes place within a quantum dot coupled to conducting leads or within one dimensional wires, often coupled to a localized magnetic moment.  In this paper, we derive a model to study the effect that the Rashba spin-orbit coupling has on the low energy behaviour of a \emph{two} dimensional system with a magnetic impurity.  The generic nature of the derived model may not be as readily applicable to the analysis of a specific experiment as those cited above, but it does have the advantage of elucidating some general features of the physical interplay between the Kondo and Rashba interactions that might be less transparent in more complicated quantum dot studies.

The model is presented in~\S\ref{sec:models} and an effective model relevent to the study of the low energy behaviour is derived.  The analysis of the effective model suggests that a more physical property of the system should be calculated in order to get a better sense of how the spin-orbit coupling influences its behaviour.  In~\S\ref{sec:resistivity}, the resistivity of the full two dimensional model is calculated and compared to Kondo's classic result~\cite{kondo}.  The combination of this result together with an analysis of the effective model leads to various conclusions and predictions which are discussed in~\S\ref{sec:discussion}.

Throughout the analysis we use the following conventions: vector quantities are represented by a boldface character and the same character without boldface denotes the vector's magnitude.  Repeated Greek indices are assumed to be summed over unless one of them is enclosed in brackets.  Finally, we work in units in which $\hbar = 1$ except when estimating the values of certain quantities in which case the precise units will be given.


\section{Model Hamiltonians}
\label{sec:models}

\subsection{Full Two Dimensional Model}

We first consider electrons confined to travel in the $xy$ plane as the result of some non-constant electric potential in the $z$ direction.  The motion of the electrons in the presence of this electric field gives rise to a coupling between the spin of the electron and its momentum, which can be expressed by the Rashba Hamiltonian~\cite{rashba,winkler}
\begin{equation}
H_R = \frac{1}{(2 \pi)^2} \int d^2 k \left\{ \frac{k^2}{2 m} \Psi_{\bs{k} \mu}^\dag \Psi_{\bs{k} \mu} + \alpha \Psi_{\bs{k} \mu}^\dag \left( k^x \sigma^y_{\mu \nu} - k^y \sigma^x_{\mu \nu} \right) \Psi_{\bs{k} \nu} \right\}.
\end{equation}
Our notation is as follows: $m$ is the effective mass of the electron, $\Psi_{\bs{k} \mu}$ is the operator that annihilates an electron of momentum $\bs{k} = (k^x, k^y)$ and spin $\mu$ and satisfies the anticommutation relationship $\{ \Psi_{\bs{k} \mu}^\dag, \Psi_{\bs{k}^\prime \nu} \} = \delta_{\mu \nu} \delta^{(2)}(\bs{k}, \bs{k}^\prime)$, $\alpha$ is the strength of the Rashba coupling which is proportional to the derivative of the electric potential in the $z$ direction, and $\sigma^i_{\mu \nu}$ are the Pauli matrices so that $\Psi_{\bs{k} \mu}^\dag \sigma^i_{\mu \nu} \Psi_{\bs{k} \nu}$ is twice the $i^{\textrm{th}}$ component of the spin of the electron.  

This Hamiltonian can easily be diagonalized to be of the form
\begin{equation}
\label{rashbaham}
H_R = \frac{1}{(2 \pi)^2} \int d^2 k \sum_{a=\pm} \eps^a_k A^\dag_{\bs{k} a} A_{\bs{k} a}.
\end{equation}
The Rashba bands are non-degenerate and take the form
\begin{equation}
\label{rashbadisp}
\eps^\pm_k = \frac{1}{2 m} \left( k \pm k_R \right)^2 - \eps_R
\end{equation}
where we have defined the Rashba momentum scale $k_R := m \alpha$ and the corresponding Rashba energy scale $\eps_R := k_R^2 / 2 m$.  The relationship between the operators that create particles in these Rashba bands and the original electron operators is 
\begin{equation}
\label{rashbaop}
A^\dag_{\bs{k} \pm} = \frac{1}{\sqrt{2}} \left( \Psi^\dag_{\bs{k} \uparrow} \pm i e^{i \theta} \Psi^\dag_{\bs{k} \downarrow} \right)
\end{equation}
where $\theta$ is the angle of the momentum $\bs{k}$ in polar coordinates and the normalization has been chosen such that $\{A_{\bs{k} a}^\dag, A_{\bs{k}^\prime b} \} = \delta_{ab} \delta^{(2)}(\bs{k}, \bs{k}^\prime)$. 

Now consider a spin-1/2 magnetic impurity\footnote{It is straightforward to generalize the results of this paper for the case of an impurity with higher spin.} at the origin, described by the spin operator $\bs{S}$. We write the interaction between this spin and the electrons using the so-called s-d exchange Hamiltonian or Kondo Hamiltonian
\begin{equation}
\label{kondofull}
H_K = \frac{V}{(2 \pi)^4} \int d^2k \, d^2k^\prime \: J_{\bs{k} \bs{k}^\prime} \mathbf{S} \cdot \Psi^\dag_{\bs{k} \mu} \frac{\boldsymbol{\sigma}_{\mu \nu}}{2} \Psi_{\bs{k}^\prime \nu},
\end{equation}
where $V$ is the two dimensional volume of the system.  This interaction can be thought of as arising from an Anderson impurity model~\cite{anderson} of a two dimensional Fermi gas interacting with a localized electronic state with strong Coulomb repulsion.  The details of this relationship (in the absence of spin-orbit coupling) are discussed in appendix~\ref{app:anderson} where the specific form for $J_{\bs{k} \bs{k}^\prime}$ is derived for various models of the hybridization between the local state and the electronic Fermi gas.

In the absence of Rashba spin-orbit coupling, the physics of the Kondo Hamiltonian is very well known~\cite{hewson}.  In the case of a ferromagnetic coupling ($J_{\bs{k} \bs{k}^\prime} < 0$) between the impurity and the electrons, the impurity essentially decouples from the electrons at low temperatures.  If the exchange interaction is antiferromagnetic, then there exists a temperature scale $T_K$, the so-called Kondo temperature, below which the magnetic impurity is screened by forming a singlet with the surrounding electrons.  Many low-energy properties can then be derived (using a variety of techniques depending on the temperature regime) and most can be shown to depend on universal scaling functions of $T / T_K$~\cite{hewson}.  For the case where the interaction occurs at a single point, $J_{\bs{k} \bs{k}^\prime} = J$ and the Kondo temperature can be expressed to lowest order in $J$ as
\begin{equation}
\label{tk0} T_K = D e^{-\frac{1}{2 J \rho_0}}
\end{equation}
where $\rho_0$ is the density of states evaluated at the Fermi energy and $D$ is an effective bandwidth.  


In semiconductor quantum dots, the value of the Kondo temperature is tunable though the maximum value is set by the size of the dot: the smaller the dot, the larger the temperature.  In this way, Kondo temperatures as high as 1 K have been obtained~\cite{goldhaber}.  The Kondo temperature of a magnetic atom on a metallic surface can span quite a wide range depending on the nature of the impurity atom and which metallic surface is used.  In the case of a Co atom on Au(111), a Kondo temperature of $T_K \approx 70$ K has been measured~\cite{surfaceexp}.

To describe a system with both a Kondo impurity and Rashba spin-orbit coupling, we can use eq.~(\ref{rashbaop}) to write the Kondo interaction compactly in the Rashba operator basis as
\begin{equation}
\label{kondoinrashba}
H_K = \frac{1}{4} \frac{V}{(2 \pi)^4} \int d^2k \, d^2k^\prime \: J_{\bs{k} \bs{k}^\prime} \sum_{a, b = \pm} S_{\bf{k} \bs{k}^\prime}^{a b} A_{\bs{k} a}^\dag A_{\bs{k}^\prime b}
\end{equation}
where 
\begin{equation}
S_{\bf{k} \bs{k}^\prime}^{a b} := i a e^{i \theta^\prime} S^- - ib e^{- i \theta} S^+ + \left( 1 - ab e^{-i ( \theta - \theta^\prime )} \right) S^z
\end{equation}
and $S^\pm := S^x \pm i S^y$ are the raising and lowering operators for the impurity spin.  On the right hand side of this equation, $a, b = \pm 1$ when the labels $a, b = \pm$ respectively.

The full two dimensional Hamiltonian, which we call the Kondo-Rashba model herein, is then $H = H_R + H_K$ where $H_R$ is given by eq.~(\ref{rashbaham}) and $H_K$ is given by eq.~(\ref{kondoinrashba}).  Given the complicated nature of the interaction between the Rashba quasi-particles with non-degenerate bands and the magnetic impurity, it is not clear at this stage whether or not the well-known physics of the Kondo effect, briefly described above, is still manifest in the Kondo-Rashba model.


\subsection{Low Energy Effective One Dimensional Model}
\label{sec:effmod}

The simplest form for the exchange interaction is to consider that it occurs at a single point.  As discussed in appendix~\ref{app:anderson}, this means that $J_{\bs{k} \bs{k}^\prime} = J$.  In this case, decomposing the operators into their different angular components reveals that only a finite number of angular modes couple to the impurity.

We first investigate the Kondo interaction in the absence of spin-oribt coupling.  With $k$ and $\theta$ being the amplitude and angle of the momentum $\bs{k}$, we can write
\begin{equation}
\label{harmexp}
\Psi_{\bs{k} \mu} = \frac{1}{\sqrt{2 \pi k}} \sum_{m=-\infty}^\infty \psi_{k \mu m} e^{i m \theta}
\end{equation}
and using the orthonormality of $(1/\sqrt{2 \pi}) e^{i m \theta}$ this implies that
\begin{equation}
\label{harmcomp}
\psi_{k \mu m} = \sqrt{\frac{k}{2 \pi}} \int_0^{2 \pi} d \theta \: e^{-i m \theta} \Psi_{\bs{k} \mu}.
\end{equation}
We have included the prefactors of $\sqrt{k}$ here so that the polar components $\psi$ of $\Psi$ behave as ordinary Fermionic operators $\{\psi^\dag_{k \mu m}, \psi_{k^\prime \nu n} \} = \delta_{\mu \nu} \delta_{m n} \delta(k - k^\prime)$.  In terms of these angular components, the Kondo interaction~(\ref{kondofull}) can be written as
\begin{equation}
\label{1dkondopsi}
H_K = \frac{J}{2} \frac{V}{(2 \pi)^3} \int_0^\infty dk \, dk^\prime \: \sqrt{k k^\prime} \boldsymbol{S} \cdot \psi^\dag_{k \mu 0}  \boldsymbol{\sigma}_{\mu \nu} \psi_{k^\prime \nu 0}.
\end{equation}
That is, the Kondo interaction is effectively a one-dimensional interaction, and all higher harmonics decouple from the impurity.

In the presence of Rashba spin-orbit coupling, it is convenient to write the Kondo interaction in terms of the operators that create and annihilate Rashba quasi-particles. First we define the following operators, which are directly related to the $m=0$ and $m=-1$ angular components of the $A_{\bs{k} \pm}$ operators respectively
\begin{eqnarray}
\label{adef1}
a_{k \pm \uparrow} & := & \sqrt{\frac{k}{2 \pi}} \int_0^{2 \pi} d \theta  \, A_\pm(k, \theta) \\
\label{adef2}
a_{k \pm \downarrow} & := & \pm i \sqrt{\frac{k}{2 \pi}} \int_0^{2 \pi} d \theta \, e^{i \theta} A_\pm(k, \theta).
\end{eqnarray}
These satisfy the usual Fermionic anti-commutation relations $\{a^\dag_{k b \mu}, a_{k^\prime c \nu} \} = \delta_{b c} \delta_{\mu \nu} \delta(k - k^\prime)$.  By combining eq.~(\ref{harmcomp}) and the inversion of eq.~(\ref{rashbaop}), one can show that
\begin{equation}
\label{psiasimp}
\psi_{k \mu 0} = \frac{1}{\sqrt{2}} \left(a_{k + \mu} + a_{k - \mu} \right).
\end{equation}
Since $\psi_{k \mu 0}$ is the only electronic mode that couples to the impurity (cf.\ eq.~(\ref{1dkondopsi})), we see that only the $m=0, -1$ modes of the Rashba quasi-particle operators couple to the impurity and all other modes can be neglected\footnote{The fact that only the -1 mode couples and not the +1 mode can be traced back to the choice of global phase in the states that diagonalize the Rashba Hamiltonian~(\ref{rashbaop}).  Of course, such a choice is arbitrary and does not effect the resulting analysis.}.  Hence, we may write the Kondo-Rashba Hamiltonian as a one dimensional model
\begin{eqnarray}
\label{1dkondorashba}
H & = & \frac{1}{(2 \pi)^2} \int_0^{\infty} dk \sum_{b = \pm} \varepsilon^b(k) a^\dag_{k b \mu} a_{k b \mu} \nonumber \\
& & + \frac{J}{4} \frac{V}{(2 \pi)^3} \int_0^\infty dk \, dk^\prime \: \sqrt{k k^\prime} \sum_{b, c = \pm} \bs{S} \cdot a^\dag_{k b \mu} \bs{\sigma}_{\mu \nu} a_{k^\prime c \nu}.
\end{eqnarray}
This has the form of a two channel Kondo model with off-diagonal couplings, where each channel has a different dispersion relation and the coupling with the impurity is the same between each of the channels.  In this expression, the Greek indices no longer refer to the spin of the electron but, instead, a combination of the spin and orbital angular degrees of freedom.  This can be seen explicitly by writing the $a_{k \pm \mu}$ in terms of the angular modes of the electrons~(\ref{harmcomp})
\begin{eqnarray}
a_{k \pm \uparrow} & = & \frac{1}{\sqrt{2}} \left( \psi_{k \uparrow 0} \mp i\psi_{k \downarrow +1} \right)\\
a_{k \pm \downarrow} & = & \frac{1}{\sqrt{2}} \left( \psi_{k \downarrow 0} \mp i\psi_{k \uparrow -1} \right).
\end{eqnarray}

Since we are interested only in the low energy properties of the Kondo-Rashba model at this stage, we can focus only on the quasi-particle excitations close to the Fermi energy.  Hence, we expand each of the dispersions to linear order about their respective Fermi momenta $k_{F \pm}$ 
\begin{equation}
\eps^\pm(k) \approx \eps_F + v_F (k - k_{F \pm}).
\end{equation}
We consider this approximation to be valid in some momenta range  $k \in (k_{F\pm} - \Lambda, k_{F\pm} + \Lambda)$ for each of the $\pm$ bands respectively, such that $k_R \ll \Lambda \ll k_F^0$.  The expressions for the Fermi momenta and velocity as derived from eq.~(\ref{rashbadisp}) are given by
\begin{eqnarray}
k_{F \pm} & = & k_F^0 \sqrt{ 1 + \frac{\eps_R}{\eps_F} } \mp k_R \\
\label{vf}
v_{F} & := & \left. \diff{\eps^\pm_k}{k} \right|_{k_{F \pm}} = v_F^0 \sqrt{1 + \frac{\eps_R}{\eps_F}}
\end{eqnarray}
where we use a superscript $0$ to indicate values for the Fermi gas in the absence of spin-orbit coupling (\textit{i.e.}~$k_F^0 = \sqrt{2 m \eps_F}$, $v_F^0 = \sqrt{2 \eps_F / m}$).  An important point to note is that the Fermi velocities are the \emph{same} for both the $+$ and the $-$ Rashba quasi-particles.

The kinetic energy of the Rashba quasi-particles (the first term in eq.~(\ref{1dkondorashba})) now reads
\begin{equation}
H_R \approx \frac{1}{(2 \pi)^2} v_F \sum_{b = \pm} \int_{-\Lambda}^\Lambda dk \: k a^\dag_{k b \mu} a_{k b \mu}
\end{equation}
where we have shifted the integration variable to be measured with respect to $k_{F\pm}$ in each of the two integrals in the summation respectively and simply relabelled the operators $a_{(k+k_{F\pm}) \pm \mu} \mapsto a_{k \pm \mu}$.  The constant $\eps_F$ part of the linearized dispersion has also been dropped.

We now similarly restrict the region of integration in the Kondo interaction (the second term in eq.~(\ref{1dkondorashba})) and expand $\sqrt{k}$ and $\sqrt{k^\prime}$ about either of $k_{F \pm}$ as appropriate.  Since all higher order terms in $k$ and $k^\prime$ are irrelevent we keep simply the constant term in each expansion.  The result is that only the combination $\sqrt{k_{F+}} a_{k+\mu} + \sqrt{k_{F-}} a_{k-\mu}$ couples to the impurity.  Hence, we define a new complete set of orthonormal operators
\begin{eqnarray}
\label{bdef}
b^\dag_{k \mu} & := & \frac{1}{\sqrt{k_{F+} + k_{F-}}} \left( \sqrt{k_{F+}} a^\dag_{k + \mu} + \sqrt{k_{F-}} a^\dag_{k - \mu} \right) \\
\tilde{b}^\dag_{k \mu} & := & \frac{1}{\sqrt{k_{F+} + k_{F-}}} \left( \sqrt{k_{F-}} a^\dag_{k + \mu} - \sqrt{k_{F+}} a^\dag_{k - \mu} \right).
\end{eqnarray}
In terms of this new basis, our effective one dimensional model is 
\begin{eqnarray}
H & = & \frac{1}{(2 \pi)^2} v_F \int_{-\Lambda}^\Lambda dk \: k \left( b^\dag_{k \mu} b_{k \mu} + \tilde{b}^\dag_{k \mu} \tilde{b}_{k \mu} \right) \nonumber \\
& & + J^\textrm{eff} \frac{V}{(2 \pi)^3} k_F^0 \int_{-\Lambda}^\Lambda dk \, dk^\prime \: \bs{S} \cdot b^\dag_{k \mu} \frac{\bs{\sigma}_{\mu \nu}}{2} b_{k^\prime \nu}.
\end{eqnarray}
If we neglect the free $\tilde{b}$ quasi-particles that do not couple to the impurity, then this Hamiltonian is of the form of a single channel, single impurity low energy Kondo model with rescaled parameters $v_F$, given by eq.~(\ref{vf}), and $J^{\textrm{eff}}$, given by
\begin{equation}
\label{scaledj} J^{\textrm{eff}} = J \frac{k_{F+} + k_{F-}}{2 k_F^0} = J \sqrt{1 + \frac{\eps_R}{\eps_F}}.
\end{equation}
As is necessary, this reduces to the original Kondo model with unscaled parameters in the limit $\eps_R \to 0$.  This is an example of the $N$ channel model studied by Simon \& Affleck~\cite{simonaffleck} where it was shown that such models can always be reduced to a one channel model provided that the Fermi velocities for all channels are the same (which is the case for the two Rashba bands as shown by eq.~(\ref{vf})).

We can interpret the change in the Fermi velocity as a change in the effective mass of the quasi-particles that interact with the impurity.  The corresponding change in the constant density of states for 2D electrons $\rho_0 = m / 2 \pi$ is then
\begin{equation}
\label{scaledrho}
\rho = \frac{\rho_0}{\sqrt{1 + \frac{\eps_R}{\eps_F}}}.
\end{equation}
Multiplying~(\ref{scaledj}) and~(\ref{scaledrho}) gives the result that 
\begin{equation}
J^{\textrm{eff}} \rho = J \rho_0.
\end{equation}  
Since the Kondo-Rashba model has been shown to reduce to the Kondo model with rescaled parameters, the Kondo temperature is the same as that of the ordinary Kondo model, eq.~(\ref{tk0}), with the above rescaled coupling
\begin{equation}
T_K = D e^{-\frac{1}{2 J^{\textrm{eff}} \rho}} = D e^{-\frac{1}{2 J \rho_0}}.
\end{equation}
That is, up to any possible changes to the energy cutoff $D$ which, in this analysis, was implemented \textit{ad hoc}, the behaviour of the Kondo-Rashba model is the same as that for the Kondo model and the value of the Kondo temperature is unchanged.



\section{Resistivity of the Two-Dimensional Model}
\label{sec:resistivity}

As we saw in the above section, any deviation in the low energy behaviour of the Kondo-Rashba model from that of the ordinary Kondo model (if there is any deviation at all) will be manifest as a change in the Kondo temperature via a change in the effective bandwidth cutoff $D$.  In this section, we bring back the full two dimensional Kondo-Rashba model that includes the full energy range of the Rashba bands up to some high energy cutoff imposed by the finite range of the impurity interaction.  In this way, band effects that were lost in the low energy linearization can be included in determining the change in the effective cutoff that appears in the expression for the Kondo temperature.  Specifically,  we calculate a physical quantity of the Kondo-Rashba model, namely, the resistivity, and compare it to the same quantity of the Kondo model, and interpret any difference as a change in this effective cutoff stemming from spin-orbit effects.  The modification to the cutoff will then be incorporated into the expression for the Kondo temperature.

As discussed in appendix~\ref{app:anderson}, a convenient choice for the finite range interaction with the localized Anderson impurity (in the absence of spin-orbit interaction) leads to a Kondo model of the form~(\ref{kondofull}) with $J_{\bs{k} \bs{k}^\prime} = J h(k) h(k^\prime)$ where
$h(k) = \Theta(\Lambda - k)$, $\Theta$ being the Heaviside step function and $\Lambda = 2 \pi / a$ being the momentum scale associated with the range of the interaction $a$.

We will calculate the conductivity using the full two dimensional Kondo-Rashba model $H = H_R + H_K$ where the terms on the right hand side are given by~(\ref{rashbaham}) and~(\ref{kondoinrashba}).  Following the original calculation of Kondo~\cite{kondo}, we use the semi-classical Boltzmann theory of transport, appropriately generalized to the present case where the two bands are no longer degenerate, as described in appendix~\ref{app:boltz}.

The resistivity is $R(T) = 1 / \sigma(T)$ where $\sigma$ is the conductivity given by equation~(\ref{newboltzcon}), specialized here to the case of the two dimensional $\pm$ Rashba bands
\begin{equation}
\label{conductivity} \sigma(T) = - \frac{e^2}{2} \int \frac{d^2 k}{(2
\pi)^2} \sum_{a = \pm} (v_k^a)^2 \tau_1^a(k) \pdiff{f^a}{\eps_k^a}.
\end{equation}
Here, $e$ is the electron charge, $\bs{v}_k^a := \nabla \eps_k^a$ is the velocity of the $a^\textrm{th}$ band, $f^a$ is the Fermi-Dirac distribution for the $a^\textrm{th}$ band, and $\tau^a_1(k)$ is the inverse of the scattering rate of the $a^\textrm{th}$ band given by eq.~(\ref{scatrate})
\begin{equation}
\label{scatrate2} 
\frac{1}{\tau^\pm_1(k)} = 2 \pi c_{\textrm{imp}} \int \frac{d^2 k^\prime}{(2 \pi)^2} \sum_{a^\prime} \left\langle
\left| T_{\bs{k} a, \bs{k}^\prime a^\prime} \right|^2
\right\rangle \left( 1 - \frac{v_{k^\prime}^{a^\prime}}{v_k^a} \cos \theta^\prime
\right) \delta(\eps_k^a - \eps_{k^\prime}^{a^\prime}).
\end{equation}
In this equation, $c_\textrm{imp}$ is the concentration of magnetic impurities (assumed to be dilute in order to neglect interactions between the impurities themselves), $\theta^\prime$ is the angle of $\bs{k}^\prime$ as measured with respect to $\bs{k}$, and $T_{\bs{k} a, \bs{k}^\prime a^\prime} = \bra{\bs{k}a} T \ket{\bs{k}^\prime a^\prime}$ is the T-matrix, \textit{i.e.}~matrix elements of the operator
\begin{equation}
\label{tmatrix}
T = H_K + H_K G_0^+ H_K + H_K G_0^+ H_K G_0^+ H_K + \cdots
\end{equation}
written in the basis of eigenstates $\ket{\bs{k} a}$ of $H_R$ with $G_0^+ = 1 / (\eps - H_R + i \eta)$ being the retarded Greens function.  The angle brackets indicate a free spin averaging over the impurity operators in the T-matrix.

\begin{figure}
\begin{center}
\psfrag{ku}{$\bs{k} +$}
\psfrag{kpu}{$\bs{k}^\prime +$}
\psfrag{qu}{$\bs{q} +$}
\psfrag{qd}{$\bs{q} -$}
\psfrag{u}{}
\psfrag{d}{}
\psfrag{+}{$+$}
\includegraphics{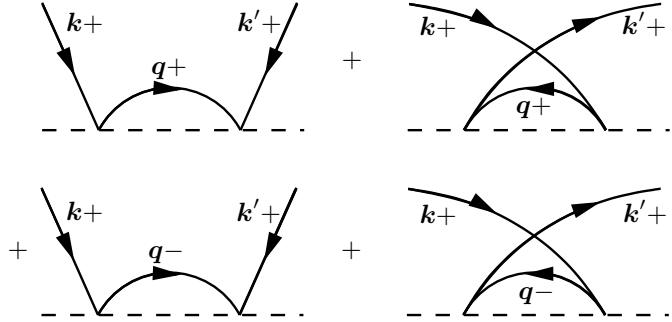}
\end{center}
\caption{\label{fig:diagtmatrix} The diagrammatic expansion of the second order contributions to the T~matrix element $T_{\bs{k}^\prime +, \bs{k} +}$ for the finite range Kondo-Rashba model.  The solid line represents the propagator for a Rashba quasi-particle, the dashed line represents the impurity, and time flows from left to right.  The internal momentum $\bs{q}$ is summed over.  After interacting with one of the Rashba quasi-particles the impurity is no longer in a spin eigenstate and so the spin states of the impurity are not labelled here.}
\end{figure}

If we consider $J \ll \eps_F$ then we can interpret eq.~(\ref{tmatrix}) as a perturbation expansion and compute it to $O(J^2)$ using the full Kondo-Rashba interaction~(\ref{kondoinrashba}), as denoted diagramatically in Fig.~\ref{fig:diagtmatrix}.  Keeping terms to \emph{third} order upon squaring and spin averaging these matrix elements results in
\begin{eqnarray}
\label{tmatrixpp} \left\langle \left| T_{\bs{k} +, \bs{k}^\prime +} \right|^2
\right\rangle & = & \left\langle \left| T_{\bs{k} -, \bs{k}^\prime
-} \right|^2 \right\rangle \nonumber \\
& = & \frac{J^2}{8} \left| h(k) h(k^\prime) \right|^2 \left( 1 +
\sin^2 \frac{\theta^\prime}{2} \right) \nonumber \\
& & \times \left[ 1 - \frac{J}{2} \left(
g^+ + g^- + \textrm{c.c.} \right) \right]
\end{eqnarray}
and
\begin{eqnarray}
\label{tmatrixpm} \left\langle \left| T_{\bs{k} -, \bs{k}^\prime +} \right|^2
\right\rangle & = & \left\langle \left| T_{\bs{k} +, \bs{k}^\prime
-} \right|^2 \right\rangle. \nonumber \\
& = & \frac{J^2}{8} \left| h(k) h(k^\prime) \right|^2 \left( 1 +
\cos^2 \frac{\theta^\prime}{2} \right) \nonumber \\
& & \times \left[ 1 - \frac{J}{2} \left(
g^+ + g^- + \textrm{c.c.} \right) \right]
\end{eqnarray}
In these equations, $\theta^\prime$ is the angle \emph{between} the two momentum vectors $\bs{k}$ and $\bs{k}^\prime$, c.c.~represents the complex conjugate of the preceding terms, and we have defined
\begin{equation}
\label{gpm} g^\pm := - \frac{V}{(2 \pi)^2} \int d^2 q \left| h(q)
\right|^2 \frac{ f(\eps_q^\pm) - \frac{1}{2} }{\eps - \eps_q^\pm + i \eta} 
\end{equation}
which comes from summing over the intermediate momentum state and is non-divergent due to the momentum cut-off coming from the finite size of the interaction via the function $h(q)$.

In order to continue with this calculation, we make some assumptions regarding the various energy scales that are present in this system.  Let $D := \Lambda^2 / 2 m$ be the energy scale associated with the high-energy cutoff that arises from the finite range of the interaction and assume that $D > \eps_F$.  Relative to the Fermi energy, the lower cutoff is approximately $\eps_F$ (the distance between the Fermi energy and the bottom of the band) and the upper cutoff is approximately $D - \eps_F$.  We have already assumed that $J \ll \eps_F$ by truncating the above perturbation expansion.  Since the spin-orbit coupling is relatively weak in all physical systems that are studied, we also assume that $\eps_R \ll \eps_F$ and $\eps_R \ll D$.  We are interested in the low temperature behaviour and can therefore assume that $T \ll \eps_F$ and $T \ll (D - \eps_F)$.  This last assumption allows us to approximate derivatives of the Fermi-Dirac distribution 
\begin{equation}
\diff{f}{\eps} \approx - \delta(\eps - \eps_F)
\end{equation}
\emph{except} in terms that diverge as $T \to 0$.  Such an approximation, as was used in Kondo's original calculation~\cite{kondo}, is valid since the dominant behaviour comes from the diverging logarithm, as will be seen.

Substituting eqs.~(\ref{tmatrixpp}) and~(\ref{tmatrixpm}) into~(\ref{scatrate2}) results in the scattering rates
\begin{equation}
\label{scatrate3} \frac{1}{\tau_1^\pm(k)} = \frac{\pi c_{\textrm{imp}} J^2 \rho_0}{8} \Omega_k^\pm \left(1 - \frac{J}{2} (g^+ + g^- + \textrm{c.c.}) \right).
\end{equation}
The function $\Omega_k^\pm$ takes different forms depending on which range of $k$ space one is in with respect to $k_R$ and $\Lambda$.  However, we are only concerned with the value of this function in the vicinity of the Fermi wave vectors $k_{F \pm}$ since $\tau_1^\pm$ is eventually multiplied by a derivative of the Fermi-Dirac distribution in eq.~(\ref{conductivity}).  Hence, we can take 
\begin{equation}
\Omega_k^\pm = 6 + \frac{k_R^2}{(k \pm k_R)^2} = 6 + \frac{\eps_R}{\eps_k^\pm + \eps_R}
\end{equation}
which is strictly valid for $2k_R < k < \Lambda - 2k_R$.

To evaluate $g^\pm$ and the integral in the expression for the conductivity, it is convenient to convert them to integrals over energy using the density of states which, for the Rashba bands, are
\begin{eqnarray}
\rho^+(\eps) & = & \rho_0 \Theta(\eps) \left[ 1 - \sqrt{\frac{\eps_R}{\eps + \eps_R}} \right]  \\
\rho^-(\eps) & = & \rho_0 \Theta(\eps + \eps_R) \left[ 1 + \sqrt{\frac{\eps_R}{\eps + \eps_R}} - \Theta(-\eps) \left( 1 - \sqrt{\frac{\eps_R}{\eps + \eps_R}} \right) \right]
\end{eqnarray}
where, as above, $\rho_0 = m / 2 \pi$ is the two dimensional density of states for a Fermi gas.  Under the approximations discussed above, inverting eq.~(\ref{scatrate3}) to leading order in $J$ and $\eps_R$ and substituting into eq.~(\ref{conductivity}) yields
\begin{equation}
\label{conductivity2} \sigma(T) = \frac{4 e^2 \left( \eps_F + \frac{5}{6} \eps_R \right)}{3 \pi^2 c_{\textrm{imp}} J^2 \rho_0} \left(1 - \frac{J}{2} \int_0^D d\eps \left( g^+ + g^- + \textrm{c.c.} \right) \diff{f}{\eps} \right).
\end{equation}
The difference in the limits of integration between the two bands has been ignored on account of the presence of $d f / d \eps$.  

However, the integrals in $g^\pm$ are evaluated by integrating by parts.  The integrands are evaluated at their respective limits and so the differences between the two bands will become manifest. Writing
\begin{equation}
g^\pm = - \int_{-\infty}^\infty d\eps^\prime \rho^\pm(\eps^\prime) \left| h(k^\pm(\eps^\prime)) \right|^2 \frac{f(\eps^\prime) - \frac{1}{2}}{\eps - \eps^\prime + i \eta},
\end{equation}
where $k^\pm(\eps^\prime)$ is the inversion of $\eps^\pm_k$, integrating by parts, substituting into eq.~(\ref{conductivity2}), and inverting to leading order in $J$ yields an expression for the resistivity
\begin{equation}
R(T) = \frac{1}{\sigma(T)} = \frac{3 \pi^2 c_{\textrm{imp}} J^2 \rho_0}{4 e^2 \left( \eps_F + \frac{5}{6} \eps_R \right)} \left[ 1 - \frac{J}{2} \left( I + \rho_0 K \right) \right].
\end{equation}
The diverging logarithm comes from the term 
\begin{equation}
I := \sum_{a, b = \pm} \rho^a(\eps_F) \int_0^{\eps_\Lambda^a} d\eps \int_0^{\eps_\Lambda^b} d \eps^\prime \ln \left| \frac{\eps - \eps^\prime}{\sqrt{\eps_F (\eps_\Lambda^b - \eps_F)}} \right| \diff{f}{\eps} \diff{f}{\eps^\prime}
\end{equation}
and all of the terms arising from the integral over the square root in the density of states have been collected in the term
\begin{eqnarray}
K := & 2 \sqrt{\frac{\eps_R}{\eps_F + \eps_R}}  \biggl(  & \atanh \sqrt{\frac{\eps_\Lambda^+ + \eps_R}{\eps_F + \eps_R}} - \atanh \sqrt{\frac{\eps_\Lambda^- + \eps_R}{\eps_F + \eps_R}} \nonumber \\ 
& &  - 2 \atanh \sqrt{\frac{\eps_R}{\eps_F + \eps_R}} \quad \biggr).
\end{eqnarray}
Again, we have used the assumptions and approximations discussed above except in the case of the log term which diverges as $T \to 0$.

We can extract the temperature dependence by making the change of variables $x^{(\prime)} = (\eps^{(\prime)} - \eps_F) / T$ in the above integral to obtain the final expression
\begin{equation}
\label{kondores}
R(T) = \frac{3 \pi^2 c_{\textrm{imp}} J^2 \rho_0}{4 e^2 \left( \eps_F + \frac{5}{6} \eps_R \right)} \left[ 1 - \frac{J}{2} \rho_0 \left( \ln \frac{T^4}{\eps_F^2 (\eps_\Lambda^+ - \eps_F)(\eps_\Lambda^- - \eps_F)} + K + C \right) \right]
\end{equation}
where $C$ is simply a finite number.  As $\eps_R \to 0$, this reduces to the classic resistivity of the Kondo model
\begin{equation}
\label{kondores0}
R^0(T) = \frac{3 \pi^2 c_{\textrm{imp}} J^2 \rho_0}{4 e^2 \eps_F } \left[ 1 - 2 J \rho_0 \left( \ln \frac{T}{D^0} + \frac{C}{4} \right) \right]
\end{equation}
with $D^0 := \sqrt{\eps_F (D - \eps_F)}$ being the UV cutoff in the absence of spin-orbit coupling.


\subsection{Interpretation of spin-orbit effects}

The analysis of the effective one dimensional model showed us that any change in the low-energy behaviour of the Kondo-Rashba system should be interpretted as a change in the cutoff.  Since we are considering the case where $\eps_R \ll \eps_F$ and $\eps_R \ll D$, we can simplify the above expression for the resistivity, eq.~(\ref{kondores}), by keeping only terms up to first order in $\eps_R/\eps_F$ and $\eps_R/D$
\begin{equation}
R(T) \approx \frac{3 \pi^2 c_{\textrm{imp}} J^2 \rho_0}{4 e^2 \left( \eps_F + \frac{5}{6} \eps_R \right)} \left[ 1 - 2 J \rho_0 \left( \ln \frac{T}{D_{\textrm{eff}}} + \frac{C}{4} \right) \right].
\end{equation}
where we have interpretted all of the changes as giving rise to an effective cutoff
\begin{equation}
\label{effcutoff} D_{\textrm{eff}} = D^0 \left( 1 + \frac{\eps_R}{\eps_F} \frac{D (D - 2 \eps_F)}{(D - \eps_F)^2} \right).
\end{equation}

Given that we had no way of determining the change in the cutoff from the low energy one dimensional model, we propose that the change in the cutoff determined from the calculation of the resistivity may give us a further clue since it incorporates the full, non-linearized bands and a physical ``real'' cutoff.  In the presence of Rashba spin-orbit coupling, the Kondo temperature would be
\begin{equation}
\label{tkfinal} T_K = D_{\textrm{eff}} e^{-\frac{1}{2 J \rho_0}} = T_K^0 \left( 1 + \frac{\eps_R}{\eps_F} \frac{D (D - 2 \eps_F)}{(D - \eps_F)^2} \right)
\end{equation}
where $T_K^0$ is the Kondo temperature using the cutoff $D^0$.

In the case where the magnetic impurity is extremely localized, $D \gg \eps_F$, this expression reduces to
\begin{equation}
\label{tkfinalreduced} T_K \approx T_K^0 \left( 1 + \frac{\eps_R}{\eps_F} \right)
\end{equation}
which is independent of the details of the cutoff.  The generality and utility of these expressions will be discussed in the following section.


\section{Discussion}
\label{sec:discussion}
The analysis presented above shows that the low energy behaviour of the Kondo model, that is, the formation of a singlet that screens the magnetic impurity, persists in the presence of a Rashba spin orbit coupling even though such an interaction breaks the $SU(2)$ symmetry so that spin states are no longer eigenstates of the Hamiltonian.  The effective one dimensional model for the Kondo-Rashba system has the form of a single impurity, single channel Kondo model with an effective mass and exchange interaction that is rescaled as a function of the strength of the spin-orbit coupling, eqs.~(\ref{scaledrho}) and~(\ref{scaledj}).  However, this rescaling combines in such away as to be cancelled out in the expression for the Kondo temperature, which is the only energy scale on which the low energy behaviour of this system depends.  However, to derive such a low energy model, one must impose an \textit{ad hoc} cutoff which may be modified in the presence of spin-orbit coupling, though it is impossible to derive such a change of the effective cutoff in this low energy formalism.

To get a better sense of how the cutoff may be modified, the resistivity of the full two dimensional Kondo-Rashba model, defined over the whole energy range up to a high energy cutoff, $D$, arising from the finite size of the interaction, was calculated and changes wrought by the Rashba coupling interpretted as giving rise to an effective cutoff, eq.~(\ref{effcutoff}).  Using such a cutoff in the expression for the Kondo temperature leads to the prediction that the presence of Rashba spin-orbit coupling will have the effect of slightly increasing or decreasing (depending on whether $D$ is greater than or less than $2\eps_F$ respectively) the Kondo temperature as a linear function of $\eps_R/\eps_F$ for small values of this dimensionless parameter.

In the calculation of the resistivity, all of the modifications that arose due to the spin-orbit coupling came from the changes in the band structure (as was manifest in the densities of state $\rho^\pm(\eps)$) and \emph{not} from the modifications to the Kondo coupling, despite the apparent complicated nature of the interaction~(\ref{kondoinrashba}).  Hence, it is not surprising that the low energy effective model, which required us to linearize the energy bands about the Fermi energy and so neglect the changes in the band structure (since both bands have the same Fermi velocity), failed to show any change in the physical behaviour of the system.

Yet, it was still necessary to impose a cutoff in the full two dimensional model by introducing a finite range to the hybridization between the impurity and the conduction electrons, the precise form of which (characterized by the envelope function $h(k)$) was chosen for simplicity.  Hence, any results that depend on the nature of the hybridization may be of questionable use.  In particular, the general expression for the Kondo temperature, eq.~(\ref{tkfinal}), depends on the energy cutoff $D$.   However, in the case of a highly localized impurity, $D \gg \eps_F$ and the universal result stated in eq.~(\ref{tkfinalreduced}) is obtained.  It is in such a localized regime that these results are expected to be of greatest utility since they are independent of the details of the cutoff mechanism.\footnote{It should be noted that the universal expression for the Kondo temperature~(\ref{tkfinalreduced}) still assumes that the hybridization between the local spin and the conduction electrons is circularly symmetric.}

\subsection{Application to Experiment}

The simple models used to derive these results were chosen so that the effects of the spin-orbit coupling would be as transparent as possible.  One may now inquire as to their applicability to real two dimensional systems that are of experimental interest.  In particular, we consider two broad possibilities: that of quantum dots constructed in semi-conductor heterostructures and that of magnetic atoms placed on a clean metallic surface.

In the case of quantum dots, we conclude that the above analysis cannot apply for two reasons.  For one, the imposition of rotational invariance in the plain of the electron gas is clearly not present in the complicated gated structures that are manufactured in the laboratory.  In our calculation of the resistivity, rotational invariance played a crucial role in maintaining the Kondo effect (as manifest by the logarithm term) in the presence of spin-orbit interactions.  To be more precise, the rotational invariance is the reason why the integrations over an intermediate $+$ and $-$ Rashba quasi-particle in the perturbative calculation of the T-matrix (as denoted diagrammatically in Fig.~\ref{fig:diagtmatrix} and represented analytically by the terms $g^+$ and $g^-$ respectively in eqs.~(\ref{tmatrixpp}) and~(\ref{tmatrixpm})) occur with equal weighting in the final T-matrix expression; the difference between the interactions involving $+$ and $-$ particles, as shown in the Hamiltonian~(\ref{kondoinrashba}), all occur with a factor of $e^{\pm i \theta}$ in the T-matrix and so integrate to zero only when you integrate over \emph{all} angles from $0$ to $2 \pi$.  This is reminiscent of the results of~\cite{zawadowski} wherein a magnetic impurity embedded in a host metal with spin-orbit interaction was studied.  It was shown that the Kondo effect is supressed only when the impurity is close to a boundary and, hence, the rotational isotropy of the system is broken.

The second reason why our analysis fails in the case of semi-conductor quantum dots is the relative strength of the energy scales.  In particular, if one assumes that the distance scale of the finite range, $a$, is of order the radius of a typical quantum dot (say $a \approx 50$ nm~\cite{goldhaber}), then this leads to an energy cutoff $D = \Lambda^2 / 2 m = 2 \pi^2 / (m a^2) \approx 0.008$ eV.  This is generally smaller than the typical values of the Fermi energy in the two dimensional electron gas of the semi-conductor heterostructure (typically $\eps_F \approx 0.03$~eV~\cite{jusserand}), completely negating our necessary assumption that $D > \eps_F$.  This suggests that a more accurate model is required to study the Kondo effect with Rashba coupling in quantum dots, some examples of which have already been studied~\cite{sun,simonin,lopez,dingdong}.

It is more likely that the results obtained in this paper are applicable to systems of magnetic atoms placed on the surface of a metal.  There, the rotational isotropy is manifest (assuming that the surface is ``clean''), and the dispersion relation of the surface conduction electrons more closely approximated by the two Rashba bands~(\ref{rashbadisp}) (as shown in various experiments, \textit{e.g.}~those described in~\cite{surfaceexp}).  Furthermore, the Fermi energy is generally much larger for the surface state electrons than in the semiconductor two dimensional electron gas ($\eps_F \approx 0.5$ eV for typical metallic surfaces) making it much more likely that the length scale over which the magnetic atom hybridizes with the conduction electrons is such that $D > \eps_F$.  Indeed, assuming that such a hybridization takes place over the observed size of the impurity atom, a radius of $a \approx 7 \, \textrm{\AA}$~\cite{surfaceexp}, then $D \approx 11 \, \textrm{eV} > \eps_F$.

As a result, one could imagine detecting a change in the Kondo temperature as a result of the Rashba spin-orbit coupling by measuring the Kondo temperature of the same magnetic atom on different metallic surfaces with different strengths of the Rashba coupling.  This is likely to be terribly difficult to detect given that the change due to spin-orbit coupling is typically very small, being linear in $\eps_R / \eps_F$ ($\eps_R / \eps_F \approx 0.003$ for the case of an Au(111) surface and immeasurable in Cu(111) and Ag(111)~\cite{surfaceexp,popovic}).  Such an effect is likely to be overwhelmed by other differences between the two systems.  A clearer signature of this effect could be seen if one were able to tune the Rashba coupling independently from the Fermi energy, something that has been demonstrated in semi-conductor heterostructures~\cite{grundler} but not for the coupling in metallic surface electrons.  Hence, for the purposes of observing this effect, it may be worth pursuing a more realistic model of a quantum dot system in such a two dimensional electron gas where one has the ability to tune the Rashba coupling.

As was demonstrated in~\cite{mw}, the presence of time reversal symmetry plays a fundamental role in the survival of the Kondo effect in the presence of spin-orbit interactions.  Another interesting question to pursue is how the presence of a magnetic field, which breaks time reversal invariance, can be incorporated into the models introduced above and how such a field modifies the results.


\section*{Acknowledgements}
I would like to thank Ian Affleck, Andrea Damascelli, Rodrigo Pereira, and Fei Zhou for useful discussions and input during the course of this work.  I would also like to acknowledge the National Science and Engineering Research Council of Canada for their support.

\appendix

\section{Derivation of the Kondo model from a finite range Anderson impurity model}
\label{app:anderson}

Here, we review the Schrieffer-Wolff transformation of an Anderson impurity model without spin-orbit coupling.  The purpose is to highlight how a finite range hybridization between the impurity state and the non-interacting conduction electrons manifests itself in the resulting Kondo interaction.  It is that Kondo interaction that is then used as input into the two dimensional Kondo-Rashba model studied in this paper.

The Kondo model used in this paper can be thought of as deriving from the Anderson impurity model~\cite{anderson} of an impurity electronic state annihilated by $d_\mu$ interacting with a sea of conduction electrons annihilated by $\Psi_{\bs{k} \mu}$.  The Hamiltonian for this model can be written as
\begin{equation}
H = \frac{1}{(2 \pi)^2} \int d^2 k \eps_k \Psi_{\bs{k} \mu}^\dag \Psi_{\bs{k} \mu}
+ \eps_d d_\mu^\dag d_\mu + U n_{d \uparrow} n_{d \downarrow} +
H_{\mathrm{hyb}}
\end{equation}
where $n_{d \mu} := d_\mu^\dag d_{(\mu)}$.  The hybridization between the impurity and conduction electrons is chosen to take the following form
\begin{equation}
H_{\mathrm{hyb}} = t \frac{V^{\frac{1}{2}}}{(2 \pi)^2} \int d^2 k \left( h(k)
d_\mu^\dag \Psi_{\bs{k} \mu} + \textrm{h.c.} \right).
\end{equation}
Here, $V$ is the two dimensional volume of the system, h.c.~denotes the Hermitian conjugate of the preceeding term and $h(k)$ is the Fourier transform of an ``envelope function'' $\tilde{h}(x)$ that describes the extent over which the impurity interacts with the conduction electrons.  This envelope function has the property that its maximum is at the origin, that it only depends on the distance from the origin (\textit{i.e.}~it is circularly symmetric), and that it decays to zero after some characteristic length scale $a$.  

One can perform a canonical transformation, the Schrieffer-Wolff transformation~\cite{schreiffer}, perturbatively in $t/U$, to obtain the s-d or Kondo model
\begin{equation}
H = \frac{V}{2 \pi} \int d^2 k \eps_k \Psi_{\bs{k} \mu}^\dag \Psi_{\bs{k} \mu} + \frac{V^2}{4 \pi^2} \int d^2 k d^2 k^\prime J_{k k^\prime} \bs{S} \cdot \Psi_{\bs{k} \mu}^\dag \frac{\bs{\sigma}_{\mu \nu}}{2} \Psi_{\bs{k}^\prime \nu}
\end{equation}
where $\bs{S} = d_\mu^\dag \frac{\bs{\sigma}_{\mu \nu}}{2} d_\nu$ is the operator of the impurity electron and 
\begin{equation}
J_{k k^\prime} = t^2 h^\ast(k) h(k^\prime) \left[
\frac{1}{U + \eps_d - \eps_{k^\prime}} + \frac{1}{\eps_k - \eps_d}
\right].
\end{equation}
A potential scattering term is also generated but is neglected in our analysis as it does not play a crucial role in determining spin-orbit effects.  If we are deep in the so-called local-moment regime where the singly and doubly occupied impurity levels are equally spaced below and above the Fermi energy (\textit{i.e.}~$\eps_k = \eps_{k^\prime} \approx \eps_F$ and $\eps_d \approx \eps_F - U/2$) then we can simplify the above to
\begin{equation}
J_{k k^\prime} \approx \frac{8 t^2}{U} h^\ast(k) h(k^\prime) \equiv J h^\ast(k) h(k^\prime).
\end{equation}

The traditional choice in most theoretical analysis is to take $h(k) = 1$, which is the Fourier transform of a delta function in position space and corresponds to having a point-like interaction.  This is the form of the envelope we will use in \S~\ref{sec:effmod}.  However, in \S~\ref{sec:resistivity}, it will be necessary to make the range of interaction finite in order to provide a more physical ultra-violet cutoff.  So, for ease of computation, we take $h(k) = \Theta(\Lambda - k)$ which corresponds to a position space envelope function
\begin{equation}
\tilde{h}(x) = \frac{\Lambda}{x} J_1(\Lambda x)
\end{equation}
with $J_\nu(z)$ being the Bessel functions of the 1st kind.  Such a function has an effective range of $a = 2 \pi / \Lambda$.


\section{Boltzmann theory of transport for non-degenerate bands}
\label{app:boltz}
In this appendix, we generalize the standard Boltzmann theory of transport, as described in~\cite{hewson}, for example, to the case where there are multiple bands, $\eps_k^a$, that are not necessarily degenerate.  Consider such a system in the presence of a weak static electric field $\bs{E}$.  This will give rise to a distribution function, $f_E^a(\bs{k})$, for each band.  The time rate of change is assumed to satisfy the relaxation hypothesis
\begin{equation}
\label{newdifff02} \diff{f_E^a(\bs{k})}{t} = -
\frac{f_E^a(\bs{k}) - f^a(k)}{\tau_1^a (\bs{k})}
\end{equation}
where $f^a(k) \equiv f(\eps^a_k)$, the Fermi-Dirac distribution.  Assuming that such time rate of change only comes about via the change in momentum $\bs{k}$ as a function of time, then we can equate the above to $(d \bs{k} / d t) \cdot \nabla f_E^a(\bs{k}) = -e \bs{E} \cdot \nabla f_E^a(\bs{k})$.  Working to first order in $E$ yields
\begin{equation}
\label{feapprox} f_E^a(\bs{k}) \approx f^a(k) + e
\tau_1^a(\bs{k}) \pdiff{f^a(k)}{\eps_k^a}
\bs{v}_k^a \cdot \bs{E}
\end{equation}
with $\bs{v}_k^a := \nabla \eps_k^a$ being the band velocity.

Substituting this into the expression for the current
\begin{equation}
\bs{J} = \sigma(T) \bs{E} = -e \int \frac{d^d k}{(2 \pi)^2}
\sum_a f_E^a (k) \bs{v}_k^a
\end{equation}
and averaging over all directions of space gives us
\begin{equation}
\label{newboltzcon} \sigma(T) = - \frac{e^2}{d} \int \frac{d^d k}{(2
\pi)^d} \sum_a (v_k^a)^2 \tau_1^a(\eps^a_k) \pdiff{f^a}{\eps_k^a}.
\end{equation}

To calculate the relaxation time, we introduce the scattering rate between the state $\ket{\bs{k} a}$ and $\ket{\bs{k}^\prime a^\prime}$ given by Fermi's golden rule
\begin{equation}
W_{\bs{k} a, \bs{k}^\prime a^\prime} = 2 \pi \left\langle
\left| T_{\bs{k} a, \bs{k}^\prime a^\prime} \right|^2
\right\rangle \delta(\eps_k^a -
\eps_{k^\prime}^{a^\prime})
\end{equation}
where $T_{\bs{k} a, \bs{k}^\prime a^\prime}$ is the T-matrix as introduced in \S~\ref{sec:resistivity}.

The rate of change of the distribution function $f_E^a(\bs{k})$ due to impurity
interactions can then be obtained by summing over all possible transitions of the state $\ket{\bs{k} a}$:
\begin{eqnarray*}
& & \diff{f_E^a(\bs{k})}{t} \\ 
& &
\begin{array}{rl} \displaystyle
	= - c_{\mathrm{imp}} \int \frac{d^d
	k^\prime}{(2 \pi)^d} \sum_{a^\prime} \biggl[ & W_{\bs{k} a,
	\bs{k}^\prime a^\prime} f_E^a(\bs{k}) \left( 1 -
	f_E^{a^\prime}(\bs{k}^\prime) \right) \\ 
	& - W_{\bs{k}^\prime
	a^\prime, \bs{k} a} f_E^{a^\prime}(\bs{k}^\prime)
	\left( 1 - f_E^a(\bs{k}) \right) \quad \biggr]
\end{array} \\
& & = - 2 \pi c_{\mathrm{imp}} \int \frac{d^d k^\prime}{(2 \pi)^d}
\sum_{a^\prime} \left\langle \left| T_{\bs{k} a,
\bs{k}^\prime a^\prime} \right|^2 \right\rangle \left(
f_E^a(\bs{k}) - f_E^{a^\prime}(\bs{k}^\prime) \right)
\delta (\eps_k^a - \eps_{k^\prime}^{a^\prime}), \\
\end{eqnarray*}
$c_{\textrm{imp}}$ being the concentration of impurities.  Equating this to $-e \bs{E} \cdot \nabla f_E^a(\bs{k})$ and using eq.~(\ref{feapprox}) results in the expression
\begin{equation}
\bs{v}_k^a \cdot \bs{E} = 2 \pi c_{\mathrm{imp}}
\tau_1^a(\bs{k}) \int \frac{d^d k^\prime}{(2 \pi)^d}
\sum_{a^\prime} \left\langle \left| T_{\bs{k} a,
\bs{k}^\prime a^\prime} \right|^2 \right\rangle \left(
\bs{v}_k^a - \bs{v}_{k^\prime}^{a^\prime} \right) \cdot
\bs{E} \, \delta(\eps_k^a - \eps_{k^\prime}^{a^\prime}).
\end{equation}

Specializing now to two dimensions, we assume that $\bs{v}_k^a$ is at most linear in $\bs{k}$ (which is true for the Rashba bands studied in this paper), that the T-matrix depends only on the angle between its two input momentum, and that the T-matrix is symmetric in that angle.  These allow us to make the following substitution within the integrand above
\begin{equation}
\bs{v}_{k^\prime}^{a^\prime} \cdot \bs{E} \to v_{k^\prime}^{a^\prime} E
\cos \theta \cos \theta^\prime =
\frac{v_{k^\prime}^{a^\prime}}{v_k^a} \bs{v}_k^a
\cdot \bs{E} \cos \theta^\prime
\end{equation}
where $\theta$ is the angle between $\bs{k}$ and $\bs{E}$ and $\theta^\prime$ is the angle between $\bs{k}$ and $\bs{k}^\prime$.  The reason this works is that the sine term that would also be included in the expression for the dot product vanishes after integrating over the angular variable.  Using this substitution we can extract an expression for the scattering rate
\begin{equation}
\label{scatrate}
\frac{1}{\tau^a_1(k)} = 2 \pi c_{\textrm{imp}} \int \frac{d^d k^\prime}{(2 \pi)^d} \sum_{a^\prime} \left\langle
\left| T_{\bs{k} a, \bs{k}^\prime a^\prime} \right|^2
\right\rangle \left( 1 -
\frac{v_{k^\prime}^{a^\prime}}{v_k^a} \cos \theta^\prime
\right) \delta(\eps_k^a - \eps_{k^\prime}^{a^\prime}).
\end{equation}
This can then be used in eq.~(\ref{newboltzcon}) to evaluate the conductivity.



\begin{thebibliography}{99}

\bibitem{kondo} J.~Kondo, Resistance Minimum in Dilute Magnetic Alloys. \textit{Prog.\ Theor.\ Phys.\ } \textbf{32}:37 (1964).

\bibitem{hewson} A.~C.~Hewson, The Kondo Problem to Heavy Fermions. CUP: Cambridge (1993).

\bibitem{goldhaber} D.~Goldhaber-Gordon, H.~Shtrikman, D.~Mahalu, D.~ Abusch-Magder, U.~Meirav, and M.~A.~Kastner, Kondo effect in a single-electron transistor. \textit{Nature} \textbf{391}:156 (1998); S.~M.~Cronenwett, T.~H.~Oosterkamp, and L.~P.~Kouwenhoven, A Tunable Kondo Effect in Quantum Dots. \textit{Science} \textbf{281}:540 (1998).

\bibitem{stm} J.~Li, W.~Schneider, R.~Berndt, and B.~Delley, Kondo Scattering Observed at a Single Magnetic Impurity. \textit{Phys.\ Rev.\ Lett.\ } \textbf{80}:2893 (1998); V.~Madhavan, W.~Chen, T.~Jamneala, M.~F.~Crommie, and N.~S.~Wingreen, Tunneling into a Single Magnetic Atom: Spectroscopic Evidence of the Kondo Resonance. \textit{Science} \textbf{280}:567 (1998).

\bibitem{rashba} E.~I.~Rashba and V.~I.~Sheka. \textit{Fiz.\ Tverd.\ Tela} (Leningrad) \textbf{Sb.~II}:162 (1959); \textbf{3}:1735 (1961). [English translation: Combinational Resonance of Zonal Electrons in Crystals Having a Zinc Blende Lattice. \textit{Sov.\ Phys.\ Solid State} \textbf{3}:1257 (1961)].

\bibitem{winkler} R.~Winkler, Spin-Orbit Coupling Effects in Two-Dimensional Electron and Hole Systems. Springer: Erlangen (2003).

\bibitem{sosemiconductorexp} T.~Hassenkam, S.~Pedersen, K.~Baklanov, A.~Kristensen, C.~B.~Sorensen, P.~E.~Lindelof, F.~G.~Pikus, and G.~E.~Pikus, Spin splitting and weak localization in (110) $\textrm{GaAs/Al}_x \textrm{G}_{1-x}$ quantum wells. \textit{Phys.\ Rev.\ B} \textbf{55}:9298 (1997) and references therein.

\bibitem{lashell} S.~LaShell, B.~A.~McDougall, and E.~Jensen, Spin Splitting of an Au(111) Surface State Band Observed with Angle Resolved Photoelectron Spectroscopy. \textit{Phys.\ Rev.\ Lett.\ } \textbf{77}:3419 (1996).

\bibitem{surfacerashba} L.~Peterson and P.~Hedeg\aa{}rd, A simple tight-binding model of spin-orbit splitting of $sp$-derived surface states. \textit{Surf.\ Sci.\ } \textbf{459}:49 (2000).

\bibitem{surfaceexp} G.~Nicolay, F.~Reinart, S.~H\"ufner, and P.~Blaha, Spin-orbit splitting of the $L$-gap surface state on Au(111) and Ag(111). \textit{Phys.\ Rev.\ B} \textbf{65}:033407 (2002).

\bibitem{mw} Y.~Meir and N.~S.~Wingreen, Spin-orbit scattering and the Kondo effect. \textit{Phys.\ Rev.\ B} \textbf{50}:4947 (1994).

\bibitem{sun} Q-F.~Sun, J.~Wang, and H.~Guo, Quantum transport theory for nanostructures with Rashba spin-orbital interaction. \textit{Phys.\ Rev.\ B} \textbf{71}:165310 (2005).

\bibitem{simonin} J.~Simonin, Kondo Quantum Dots and the Novel Kondo-Doublet Interaction. \textit{Phys.\ Rev.\ Lett.} \textbf{97}:266804 (2006).

\bibitem{lopez} R.~L\'opez, D.~S\'anchez, and L.~Serra, From Coulomb blockade to the Kondo regime in a rashba dot.  \textit{Phys.\ Rev.\ B} \textbf{76}:035307 (2007)

\bibitem{dingdong} G-H.~Ding and B.~Dong, Spin-orbit coupling effect on the persistent currents in mesoscopic ring with an Anderson impurity. [arXiv:0704.0319] (2007).

\bibitem{simonaffleck} P.~Simon and I.~Affleck, Kondo screening cloud effects in mesoscopic devices.  \textit{Phys.\ Rev.\ B} \textbf{68}:115304 (2003).

\bibitem{anderson} P.~W.~Anderson and A.~M.~Clogston. Bull.\ Am.\ Phys.\ Soc.\ \textbf{6}:124 (1961);  P.~W.~Anderson, Localized Magnetic States in Metals.  \textit{Phys.\ Rev.} \textbf{124}:41 (1961).

\bibitem{jusserand} B.~Jusserand, D.~Richards, H.~Peric, and B.~Etienne, Zero-magnetic-field spin splitting in the GaAs conduction band from Raman scattering on modulation-doped quantum wells. \textit{Phys.\ Rev.\ Lett.\ } \textbf{69}:848 (1992); B.~Jusserand, D.~Richards, G.~Allan, C.~Priester, and B.~Etienne, Spin orientation at semiconductor heterointerfaces.  \textit{Phys.\ Rev.\ B} \textbf{51}:4707 (1995).

\bibitem{popovic} D.~Popovi\'c, F.~Reinert, S.~ H\"ufner, M.~Springborg, H.~Cercellier, Y.~Fagot-Revurat, B~Kierren, and D.~Malterre, High-resolution photoemission on Ag/Au(111): Spin-orbit splitting and electronic localization of the surface state.  \textit{Phys.\ Rev.\ B} \textbf{72}:045419 (2005).

\bibitem{schreiffer} J.~R.~Schrieffer and P.~A.~Wolff, Relation between the Anderson and Kondo Hamiltonians. \textit{Phys.\ Rev.\ } \textbf{149}:491 (1966).

\bibitem{zawadowski} O.~\'Ujs\'aghy and A.~Zawadowski, Spin-orbit-induced magnetic anisotropy for impurities in metallic samples. I. Surface anisotropy. \textit{Phys.\ Rev.\ B} \textbf{57}:11598 (1998); Spin-orbit-induced magnetic anisotropy for impurities in metallic samples. II. Finite-size dependence in the Kondo resistivity.  \textit{Phys.\ Rev.\ B} \textbf{57}:11609 (1998).

\bibitem{grundler} D.~Grundler, Large Rashba Splitting in InAs Quantum Wells due to Electron Wave Function Penetration into the Barrier Layers.  \textit{Phys.\ Rev.\ Lett.\ } \textbf{84}:6074 (2000).

\end{thebibliography}
\end{document}